\documentclass[showpacs,aps,pre,twocolumn,amsmath,amssymb,epsfig]{revtex4-1}

\usepackage{graphicx}
\usepackage{epsfig}
\usepackage{dcolumn}
\usepackage{bm}
\newcommand     {\beq}[1]         { \begin{equation} #1 \end{equation} }
\newcommand     {\beqa}[1]        { \begin{eqnarray} #1 \end{eqnarray} }

\begin{document}

\title{Size scaling of failure  strength with fat-tailed disorder in a fiber bundle model}

 \author{Vikt\'oria K\'ad\'ar}
 \author{Zsuzsa Danku}
  \author{Ferenc Kun}
 \email{Corresponding author: ferenc.kun@science.unideb.hu}
  \affiliation{Department of Theoretical Physics, University of Debrecen,
 P.O. Box 5, H-4010 Debrecen, Hungary}

\begin{abstract}
We investigate the size scaling of the macroscopic fracture strength of
heterogeneous materials when microscopic disorder is controlled 
by fat-tailed distributions. We consider a fiber bundle model where the strength
of single fibers is described by a power law distribution over a finite range.
Tuning the amount of disorder by varying the power law exponent and the upper cutoff 
of fibers' strength, in the limit of equal load sharing an astonishing size effect 
is revealed: For small system sizes the bundle strength increases with the number 
of fibers and the usual decreasing size effect of heterogeneous materials is only
restored beyond a characteristic size. We show analytically that the extreme order 
statistics of fibers' strength is responsible for this peculiar behavior. Analyzing 
the results of computer simulations we deduce a scaling form which describes the 
dependence of the macroscopic strength of fiber bundles on the parameters of microscopic
disorder over the entire range of system sizes.
 \end{abstract}

\maketitle

\section{Introduction}
The disorder of materials plays a crucial role in their fracture processes
under mechanical loading. Strength fluctuations of local material elements 
can lead to crack nucleation at low loads reducing the failure strength compared
to homogeneous materials \cite{alava_statistical_2006,alava_role_2008,alava_size_2009}. 
Additionally, disorder gives rise to sample-to-sample
fluctuations of fracture strength with an average value which depends on the 
system size \cite{alava_size_2009}. 
This so-called size effect of the fracture strength of materials 
has a great importance for applications: on the one hand it has to be taken 
into account in engineering design of large scale construction, and on the other 
hand, it controls how results of laboratory measurements can be scaled up to
real constructions and to the scale of geological phenomena 
\cite{alava_statistical_2006,alava_role_2008,alava_size_2009,yamamoto_PhysRevE.83.066108}. 

The statistics of fracture strength and the associated 
size effect are usually described by 
extreme value theory \cite{galambos_asymptotic_1978} which relates the 
macroscopic strength of materials to the 
statistics of weakest microscopic regions 
\cite{weibull_statistical_1939,bazant_fracture_1997,alava_size_2009}. 
Weibull gave the first quantitative explanation
of the statistical size effect formulating the weakest link idea, namely, the volume 
element of the weakest flaw drives the failure of the entire system, and he determined
the probability distribution of failure strength of macroscopic samples.
In order to investigate how the enhanced stress around cracks and the interaction 
between cracks affect the strength,
stochastic lattice models of materials have been widely used 
\cite{herrmann_fracture_1989,hansen_rupture_1989,batrouni_fracture_1998,
nukala_percolation_2004,alava_size_2009,alava_role_2008}. 
In these models
disorder is represented either by random dilution of regular lattices or by the
random strength of cohesive elements. Such model calculations confirmed that extreme value 
statistics describes the distribution of macroscopic strength, however, the general validity of the 
Weibull distribution has been questioned although it is widely used in engineering
design \cite{zapperi_PhysRevApplied.2.034008}.

To study the size scaling of fracture strength the fiber bundle model (FBM) 
provides also an adequate framework 
\cite{de_arcangelis_scaling_1989,andersen_tricritical_1997,hansen2015fiber,
kloster_burst_1997,kun_extensions_2006,
hidalgo_avalanche_2009}. In FBMs the sample is discretized in terms of parallel fibers
where controlling the mechanical response, strength and interaction of fibers various
types of mechanical responses can be represented.
Additionally, FBMs are simple enough to obtain analytic solutions for the most 
important quantities of interest.
For the redistribution of load after fiber breaking two limiting cases are very useful
to study, i.e.\ the equal (ELS) and local (LLS) load sharing: 
under ELS the excess load after failure events is equally shared by all the intact fibers, and hence,
the stress field remains homogeneous all over the loading process. 
For LLS the load dropped by the broken fiber is equally shared by the intact
elements of its local neighborhood resulting in a high stress concentration
along broken clusters of fibers.

For ELS analytic calculations have revealed 
\cite{smith_asymptotic_1982,mccartney_statistical_1983,hansen2015fiber} that in the limit of large bundle 
size $N$ the average values $\left<\sigma_c\right>$ and $\left<\varepsilon_c\right>$ of the 
fracture stress $\sigma_c$ and strain $\varepsilon_c$ converge to finite values according
to a power law functional form
\beqa{
\left<\sigma_c\right>(N) &=& \sigma_c(\infty) + AN^{-\alpha}, \label{eq:els_sigc}\\
\left<\varepsilon_c\right>(N) &=& \varepsilon_c(\infty) + BN^{-\alpha}.  \label{eq:els_epsc}
}
Here $\sigma_c(\infty)$ and $\varepsilon_c(\infty)$ denote the asymptotic bundle strength.
The scaling exponent $\alpha$ has the value $\alpha=2/3$ which proved to be universal
for a broad class of disorder distributions, while the multiplication factors $A$ and $B$
depend on the specific type of disorder 
\cite{smith_asymptotic_1982,mccartney_statistical_1983,hansen2015fiber}. 

For LLS numerical calculations showed that the macroscopic strength of bundles, where the 
strength distribution of single fibers expands to zero, diminishes as the system size $N$
increases. The convergence to zero strength is logarithmically slow with the functional form 
\beq{
\left<\sigma_c\right>(N) \sim 1/(\ln{N})^{\beta},
}
where the exponent $\beta$ was found to depend on the precise range of load sharing 
\cite{harlow_pure_1985,hansen_burst_1994,hidalgo_fracture_2002,dill-langer_size_2003,
yewande_time_2003,bazant_activation_2007,lehmann_breakdown_2010,pradhan_failure_2010}.
For some modalities of stress transfer an even slower asymptotic convergence 
$\left<\sigma_c\right>(N)\sim 1/\ln (\ln N)$ to zero
strength was found as e.g.\ for hierarchical load transfer 
\cite{newman_failure_1991}. The effect of the range of load sharing on the fracture 
strength of fiber bundles has been studied at moderate amount of disorders where the 
strength of single fibers is typically sampled from a uniform, exponential, or Weibull distribution.
However, the precise amount of disorder may have a strong effect on the size scaling 
of fracture strength, which has not been explored.

In the present paper we investigate the effect of the amount of micro-scale 
disorder on the size scaling of the macroscopic strength of equal load sharing 
fiber bundles focusing on the limiting case of extremely high disorder. 
We consider a power law distribution of fibers' strength over a finite range 
where the amount of disorder
can be controlled by the exponent and by the upper cutoff of the strength values. 
As the most remarkable result, our study revealed that in a range of parameters the bundle 
strength increases with the system size. The usual decreasing behavior sets in only 
beyond a characteristic system size which depends on the amount of disorder. 
We give a quantitative explanation of these novel findings in terms of extreme order 
theory. 
The results may have potential applications for materials' design.

\section{Fiber bundle model with fat-tailed disorder}
In our model we consider a bundle of $N$ parallel fibers, which are 
assumed to have a perfectly brittle behavior, i.e.\ 
they exhibit a linearly elastic response with a Young modulus $E$ up to breaking 
at a threshold load $\sigma_{th}$. The Young modulus is assumed to be 
constant $E=1$ such that the disorder of the material is solely represented 
by the randomness of the breaking threshold $\sigma_{th}$:
to each fiber a threshold value is assigned $\sigma_{th}^i$, $i=1,\ldots , N$ 
sampled from the probability density $p(\sigma_{th})$. The amount of disorder 
in the system can be controlled by varying the range $\sigma_{th}^{min}\leq \sigma_{th} \leq 
\sigma_{th}^{max}$ of strength values and by the functional form of $p(\sigma_{th})$. 

In order to explore the effect of extremely high disorder, we consider
a power law distribution of threshold values over a finite range.  
The probability density function is written in the form
\begin{eqnarray}
\ p(\sigma_{th}) = \left\{  
\begin{array}{ccc}
0, &  \sigma_{th}< \sigma_{th}^{min},\\ [2mm]
\ A \sigma_{th}^{-(1+\mu)}, &  \sigma_{th}^{min}\leq\sigma_{th}\leq\sigma_{th}^{max}, \\ [2mm]
0, &  \sigma_{th}^{max}<\sigma_{th}, \\
\end{array}
\right.            
\label{surusegfv}
\end{eqnarray}
where the lower bound of thresholds $\sigma_{th}^{min}$ was fixed to $\sigma_{th}^{min}=1$.
The amount of disorder is controlled by varying the exponent $\mu$ of the power law and 
the upper bound $\sigma_{th}^{max}$ of the breaking thresholds, while all other parameters
are fixed.
The value of the exponent is varied over the interval $0\leq \mu \leq 1$ because 
in the limiting case of an infinite upper bound $\sigma_{th}^{max}\to \infty$ at these
$\mu$ values the disorder is so high that the thresholds do not have a finite average.
For finite values of $\sigma_{th}^{max}$, of course, the average $\left<\sigma_{th}\right>$ 
is always finite, however, the specific values of $\sigma_{th}^{max}$ and $\mu$ have 
a very strong 
effect on the behavior of the system both on the macro- and micro-scales.

After normalizing the pobability density $p(\sigma_{th})$ the cummulative distribution 
function $P(\sigma_{th})$ can be cast into the form
\begin{equation}
\label{eq:distrib}
P(\sigma_{th})= \left\{  
\begin{array}{cc}
0 & \sigma_{th}<\sigma_{th}^{min},\\[1mm]
{\displaystyle \frac{\sigma_{th}^{-\mu} - (\sigma_{th}^{min})^{-\mu}}{(\sigma_{th}^{max})^{-\mu} - 
(\sigma_{th}^{min})^{-\mu}}}, &  \sigma_{th}^{min}\leq \sigma_{th}\leq \sigma_{th}^{max},\\ [1mm]
1 & \sigma_{th}^{max}<\sigma_{th}.\\
\end{array}
\right.
\end{equation}
\begin{figure}
\begin{center}
\epsfig{file=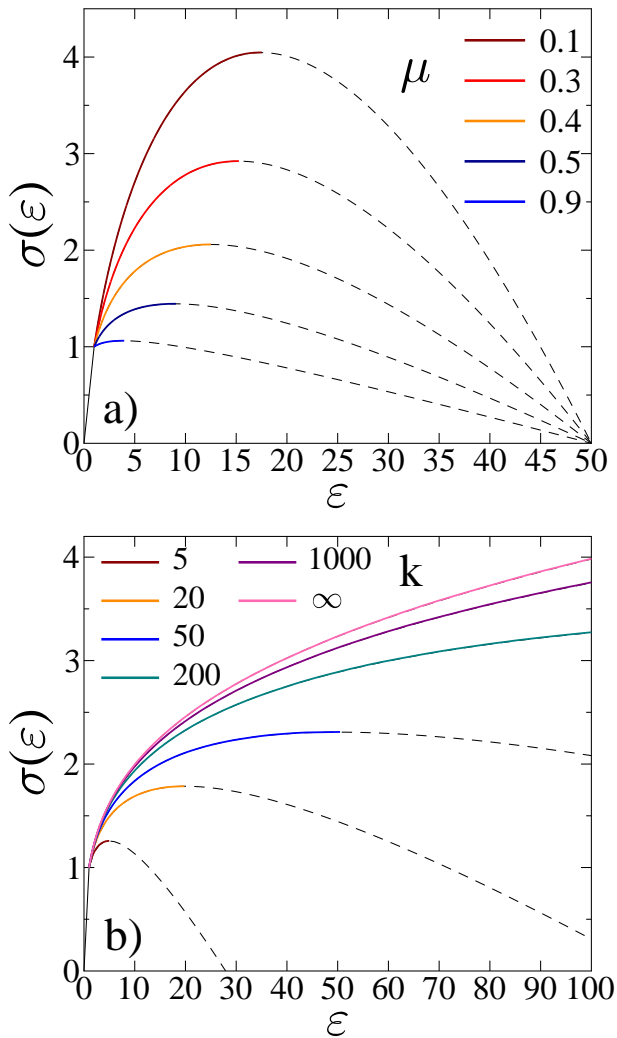,bbllx=10,bblly=10,bburx=360,bbury=590,
width=8.5cm}
\caption{\small (Color online) $(a)$ The macroscopic response $\sigma(\varepsilon)$ of the 
system for the same upper cutoff $\varepsilon_{max}=50$ 
varying the value of the exponent $\mu$. 
$(b)$ Constitutive curves for a fixed $\mu=0.7$ exponent varying 
the upper cutoff $\varepsilon_{max}$ with the multiplication factor $k$.
Approaching the phase boundary in both cases the system becomes 
more and more brittle.
The curve on the top of $(b)$ corresponds to the case $\varepsilon_{max}\to \infty$.
The dashed lines represent the full analytical curves of Eq.\ 
(\ref{sigma_th_epsilon_analitikus}) while the colored dots show the results 
of stress controlled simulations.
}   
\label{epsilon_sigma_k} 
\end{center}
\end{figure}

The macroscopic response of the bundle is characterized by the constitutive equation
$\sigma(\varepsilon)$. Assuming equal load sharing $\sigma(\varepsilon)$ can be cast 
in the general form $\sigma(\varepsilon)=E\varepsilon[1-P(E\varepsilon)]$,
where the term $1-P(E\varepsilon)$ provides the fraction of intact fibers at strain 
$\varepsilon$, which all keep the same load $E\varepsilon$ 
\cite{hansen2015fiber,kun_extensions_2006,hidalgo_avalanche_2009}.
Substituting the cummulative distribution function $P(x)$ from Eq.\ (\ref{eq:distrib})
we arrive at
\begin{equation}
\label{sigma_th_epsilon_analitikus}
\sigma(\varepsilon)= \left\{
\begin{array}{cc}
\varepsilon, & 0\leq \varepsilon \leq \varepsilon_{min},\\
\displaystyle{\frac{\varepsilon \big(\varepsilon^{-\mu}- \varepsilon_{max}^{-\mu}\big)}{\varepsilon_{min}^{-\mu}-\varepsilon_{max}^{
-\mu}}}, & \varepsilon_{min}\leq \varepsilon\leq \varepsilon_{max},\\
0,& \varepsilon_{max}<\varepsilon,\\
\end{array}
\right.
\end{equation}
where for clarity the notation $\varepsilon_{min}=\sigma_{th}^{min}/E$, 
$\varepsilon_{max}=\sigma_{th}^{max}/E$ was introduced with $E=1$.
The macroscopic constitutive response of the system is illustrated in Fig.\ 
\ref{epsilon_sigma_k}. Up to the lower bound $\varepsilon_{min}$ a perfectly linearly
elastic response is obtained since no breaking can occur. When the fibers start to break 
above $\varepsilon_{min}$ the constitutive curve $\sigma(\varepsilon)$ becomes 
non-linear and beyond the maximum it decreases to zero as all fibers break gradually.

The fracture strength of the bundle is defined by the value $\sigma_c$ of the maximum
of the constitutive curve and by its position $\varepsilon_c$ called critical stress
and strain, respectively. Under stress controlled loading exceeding the value of $\sigma_c$
the bundle rapidly undergoes global failure so that the entire $\sigma(\varepsilon)$ curve 
can only be realized under strain controlled loading. Of course, the critical strain depends 
on the degree of disorder characterized by $\mu$ and $\varepsilon_{max}$
\begin{equation}
\varepsilon_c = \varepsilon_{max}(1-\mu)^{1/\mu},
\label{eq:crit_strain}
\end{equation}
while the critical stress $\sigma_c$ depends on the lower cutoff $\varepsilon_{min}$ 
as well
\begin{equation}
\sigma_c= \frac{\mu(1-\mu)^{1/\mu-1}
\varepsilon_{max}^{1-\mu}}{\varepsilon_{min}^{-\mu}-\varepsilon_{max}^{-\mu}}.
\label{eq:sigma_c}
\end{equation}

It is a very interesting feature of the system that if the threshold distribution 
is too narrow already the first fiber breaking can trigger a catastrophic avalanche
of fiber breaking giving rise to global failure. 
This occurs when the position of the maximum of the constitutive curve $\varepsilon_c$ 
coincides with the lower bound $\varepsilon_{min}$. Keeping $\varepsilon_{min}$ fixed a threshold 
value $\varepsilon_c^{max}$ of the upper bound $\varepsilon_{max}$ can be derived as 
\beq{
\varepsilon_{max}^c=\frac{\varepsilon_{min}}{(1-\mu)^{1/\mu}}.
\label{eq:phase_boundary}
}
It follows that those bundles where $\varepsilon_{max} < \varepsilon_{max}^c$ holds, behave 
in a completely brittle way, i.e.\ macroscopic failure occurs right after the linear regime of 
$\sigma(\varepsilon)$ at the instant of the first fiber breaking. 
However, in the parameter regime $\varepsilon_{max} > \varepsilon_{max}^c$
a quasi-brittle response is obtained where macroscopic failure is preceded by breaking avalanches.
\begin{figure}
\begin{center}
\epsfig{bbllx=30,bblly=40,bburx=380,bbury=330, file=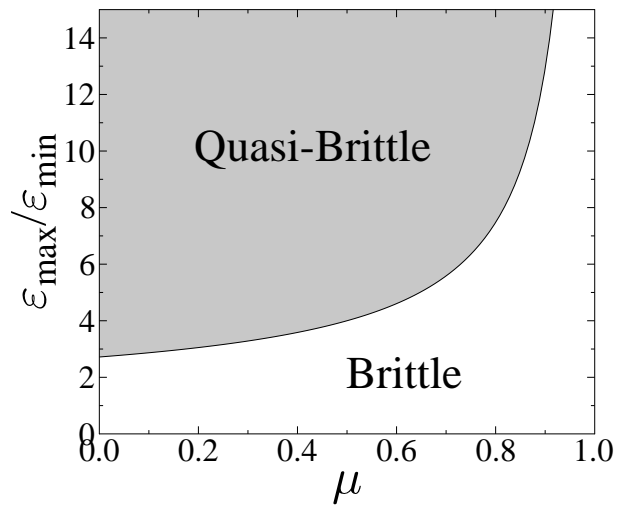, width=8.3cm}
  \caption{ Phase diagram of the system. The phase boundary separating the 
  brittle and quasi-brittle macroscopic response is given by Eq.\ (\ref{eq:phase_boundary}).
  Note that for $\mu\geq 1$ the bundle is always in the brittle phase.
   \label{fig:crack}}
\end{center}
\end{figure}
It can be observed that as the exponent $\mu$ approaches 1 from below the value of  
$\varepsilon_{max}^c$ diverges so that the regime $\mu\geq 1$ is always brittle.
The phase diagram of the system is illustrated in Fig.\ \ref{fig:crack}.

In the following we analyze how the amount of disorder affects the macroscopic 
strength of finite bundles in the quasi-brittle phase. For clarity, in these calculations
the upper cutoff $\varepsilon_{max}$ will be expressed in terms 
of $\varepsilon_{max}^c$ as $\varepsilon_{max}=k\varepsilon_{max}^c$, 
where the multiplication factor $k$ can take any value in the range $k\geq 1$.

\section{Fracture strength of finite bundles}
The fracture strength characterized by the critical strain $\varepsilon_c$ and stress 
$\sigma_c$ have been obtained analytically in Eqs.\ (\ref{eq:crit_strain},\ref{eq:sigma_c})
as function of the parameters of the model $\mu, \varepsilon_{min}$, and $\varepsilon_{max}$.
These analytical calculations assume an infinite system size so that $\varepsilon_c$ and $\sigma_c$
are the $N\to \infty$ asymptotic strength of the bundle. 
In order to reveal how the finite size of the bundle $N$ affects the average value of the 
critical strain $\left<\varepsilon_c\right>$ and stress $\left<\sigma_c\right>$
we performed computer simulations
varying the number of fibers $N$ over six orders of magnitude.
Stress controlled loading of the bundles was performed until the catastrophic avalanche
gave rise to global failure. The critical values $\varepsilon_c$ and $\sigma_c$
were determined as the strain and stress of the last stable configuration of the system
(see also Fig.\ \ref{epsilon_sigma_k}).

\begin{figure}
\begin{center}
\epsfig{bbllx=30,bblly=30,bburx=380,bbury=640, file=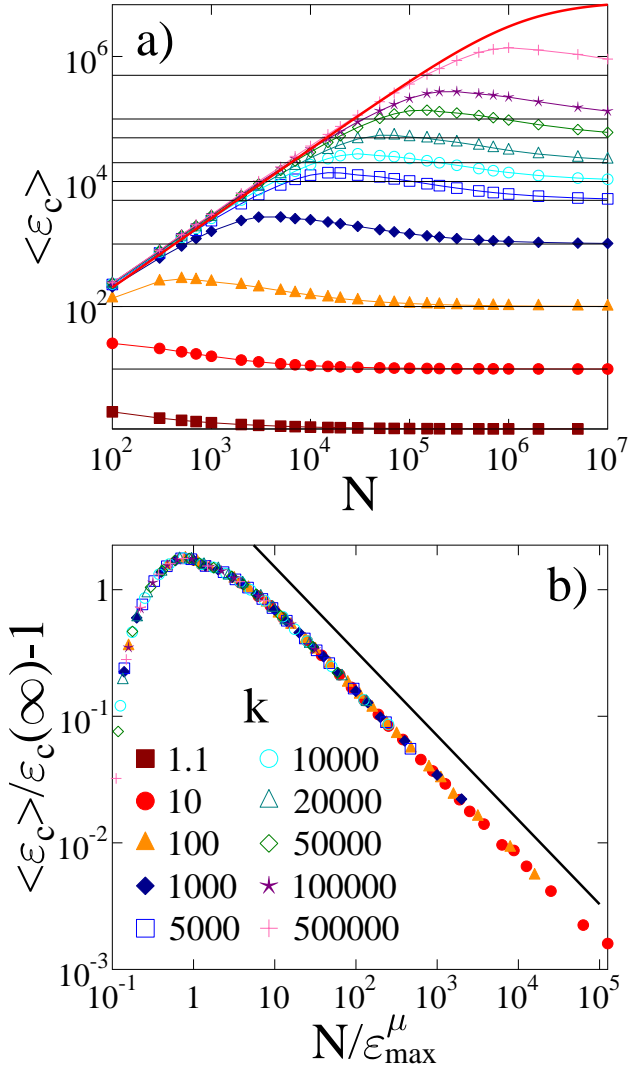, width=8.3cm}
  \caption{(Color online) $(a)$ The average value of the critical strain 
  $\left<\varepsilon_c\right>$ as a function of the bundle size $N$
  for several values of the upper cutoff $\varepsilon_{max}$ of the strength of single fibers.
  The horizontal lines represent the corresponding asymptotic strength obtained from 
  Eq.\ (\ref{eq:crit_strain}). The value of the exponent $\mu$ is fixed to $\mu=0.8$.
  The upper cutoff $\varepsilon_{max}$ is parametrized by $k$ such that the legend 
  is the same as in Fig.\ \ref{fig:eps_c}$(b)$. The red bold line
  gives the analytic curve of  Eq.\ (\ref{eq:extrem_order}). 
  $(b)$ Scaling plot of the data presented in $(a)$. After rescaling with the asymptotic 
  strength $\varepsilon_c(\infty)$ we subtracted 1 from the result in order to demonstrate 
  the asymptotic power law behavior. The straight line represents a 
  power law of exponent $-2/3$.
   \label{fig:eps_c}}
\end{center}
\end{figure}
\begin{figure}
\begin{center}
\epsfig{bbllx=25,bblly=25,bburx=360,bbury=330, file=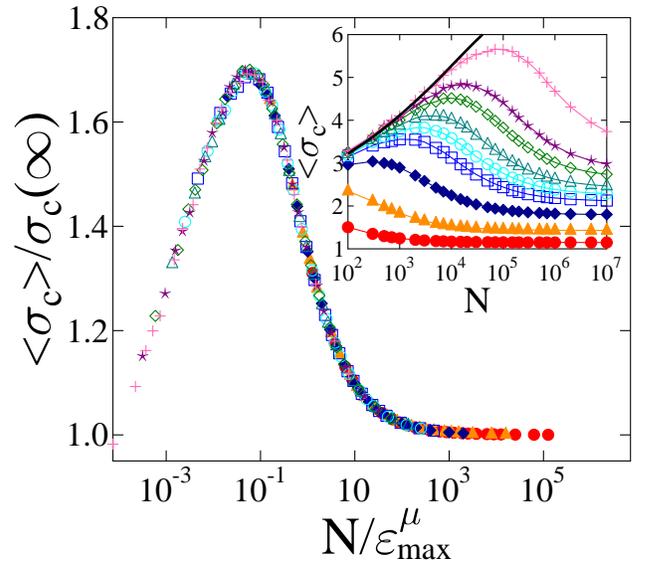, width=8.3cm}
  \caption{(Color online) Inset: the average fracture stress $\left<\sigma_c\right>(N)$ as 
  a function of the number of fibers $N$ for the same values of the upper cutoff
  as in Fig.\ \ref{fig:eps_c} using also the same legend. The bold line represents the curve
  of Eq.\ (\ref{eq:sigc_epsc_rel}).
  Main panel: Rescaling the two axis of the inset
  the curves obtained at different cutoff values can be collapsed on a master curve. 
  \label{fig:sigc_scaling}}
\end{center}
\end{figure}

It can be seen in Fig.\ \ref{fig:eps_c}$(a)$ that for low values of the upper cutoff 
$\varepsilon_{max}$ of the strength of single fibers the average bundle strength 
$\left<\varepsilon_c\right>$ monotonically decreases with increasing system size 
as it is expected. 
However, above a certain value of $\varepsilon_{max}$ the macroscopic strength 
has an astonishing unceasing regime for small system sizes so that the usual 
decreasing behavior of strength 
is restored only above a characteristic system size $N_c$. 
The horizontal lines in the figure show
that in the limit of large $N$ the average strength $\left<\varepsilon_c\right>$ converges 
to the analytic asymptotic value of Eq.\ (\ref{eq:crit_strain}). 
Note that the characteristic system size $N_c$, which separates the increasing and decreasing 
regimes of macroscopic strength, is an increasing function of $\varepsilon_{max}$.
The same qualitative behavior is observed for the critical stress $\left<\sigma_c\right>$
in the inset of Fig.\ \ref{fig:sigc_scaling}, which clearly demonstrates that the fracture strength 
of the fiber bundle increases
for small system sizes when the amount of disorder is sufficiently high. The position of the 
maximum of $\left<\sigma_c\right>$ coincides with that of $\left<\varepsilon_c\right>$. 

Figure \ref{fig:eps_c}$(b)$ and Fig.\ \ref{fig:sigc_scaling} demonstrate that rescaling 
the two axis of Fig.\ \ref{fig:eps_c}$(a)$ and of the inset of Fig.\ \ref{fig:sigc_scaling}
the curves of different cutoff values can be collapsed on a master curve. On the horizontal axis
the number of fibers is rescaled with $\varepsilon_{max}^{\mu}$, while along the vertical axis
the rescaling is performed with the corresponding asymptotic strength $\varepsilon_c(\infty)$ and 
$\sigma_c(\infty)$ in the two figures. The good quality collapse implies the scaling structures
\beqa{
\left<\varepsilon_c\right>(N,\varepsilon_{max}) &=& 
\varepsilon_c(\infty) \Phi(N/\varepsilon_{max}^{\mu}),  \\ [4mm]
\left<\sigma_c\right>(N,\varepsilon_{max}) &=& 
\sigma_c(\infty)\Psi(N/\varepsilon_{max}^{\mu}),
}
where $\Phi(x)$ and $\Psi(x)$ denote the scaling functions. 
The structure of the scaling functions $\Phi(x)$ and $\Psi(x)$ has the consequence 
that the characteristic system size $N_c$ depends on the parameters as 
\beq{
N_c\sim \varepsilon_{max}^{\mu}.
}

In Figure \ref{fig:eps_c}$(b)$ unity is subtracted from the scaling function $\Phi(x)$
which results in an asymptotic power law decrease. This behavior implies the validity 
of the functional form 
\beq{
\Phi(x) \approx 1+Cx^{-\alpha}
\label{eq:phi}
}
for $x>1$. The value of the exponent was found to be $\alpha=2/3$ which 
is consistent with the generic behavior Eq.\ (\ref{eq:els_sigc}) 
of the strength of ELS bundles. 
Note that the scaling function $\Psi(x)$ of $\left<\sigma_c\right>$
has the same features in Fig.\ \ref{fig:sigc_scaling} as $\Phi(x)$ so 
that $\Psi(x)$ can also be described by Eq.\ (\ref{eq:phi}).

\section{Extreme order statistics}
The peculiar size scaling of macroscopic strength obtained in our simulations
is the direct consequence of the fat tailed strength distribution of single fibers.
The main effect of the fat tail is that even for a small system size $N$ 
the probability to have strong fibers in the bundle can be relatively high.
Under equal load sharing conditions all fibers keep the same load so that fibers 
break in the increasing order of their breaking thresholds.
Our assumption is that for those system sizes $N$ which are along the increasing 
regime of $\left<\varepsilon_c\right>(N)$ and $\left<\sigma_c\right>(N)$ 
the strongest fibers are so strong that
a few of them or even a single one is able to keep the entire load that has been 
put on the bundle. 
It follows that the average macroscopic strength 
$\left<\varepsilon_c\right>(N)$ should be determined by the average strength of 
the strongest fiber $\left<\varepsilon_{th}^{max}\right>_N$. The average 
of the largest value of a set of $N$ random numbers sampled from the same probability 
distribution can be obtained analytically as
\beq{
\left<\varepsilon_c\right>(N)=\left<\varepsilon_{th}^{max}\right>_N = P^{-1}\left(1-\frac{1}{N+1}\right),
}
where $P$ denotes the cumulative distribution of failure thresholds.
Substituting $P$ from Eq.\ (\ref{eq:distrib}), the above expression 
yields for the macroscopic strength
\beqa{
\left<\varepsilon_{th}^{max}\right>_N &=& \left[\left( (\varepsilon_{th}^{max})^{-\mu}-
(\varepsilon_{th}^{min})^{-\mu} \right) \left(1-\frac{1}{N+1}\right)+ \right. \nonumber \\
&+& \left. (\varepsilon_{th}^{min})^{-\mu}\right]^{-1/\mu}.
\label{eq:extrem_order}
}
It can be observed in Fig.\ \ref{fig:eps_c} that Eq.\ (\ref{eq:extrem_order}) 
provides a high quality description of the increasing macroscopic fracture strength 
with the system size.
Deviations occur only around the characteristic system size $N_c$
where the curve of $\left<\varepsilon_{th}^{max}\right>_N$ saturates since the average 
of the largest cannot exceed the value of the upper cutoff of fibers' strength 
$\varepsilon_{max}$. Note that for large upper cutoffs $\varepsilon_{max}\to \infty$
the above expression Eq.\ (\ref{eq:extrem_order}) predicts a power law increase 
of the fracture strain with the system size \cite{danku_disorder_2016}
\beq{
\left<\varepsilon_c\right>(N) \sim N^{1/\mu}.
}
For the fracture stress $\left<\sigma_c\right>(N)$ it follows from the above 
arguments that along the increasing branch in Fig.\ \ref{fig:sigc_scaling}
the relation holds
\beq{
\left<\sigma_c\right>(N) = \frac{E\left<\varepsilon_c\right>(N)}{N}.
\label{eq:sigc_epsc_rel}
}
This explains the orders of magnitude difference of $\left<\varepsilon_c\right>(N)$ and
$\left<\sigma_c\right>(N)$ in Figs.\ \ref{fig:eps_c} and \ref{fig:sigc_scaling}
and the slower increase of the fracture stress
\beq{
\left<\sigma_c\right>(N) \sim N^{1/\mu-1}.
\label{eq:sigc_extreme}
}
This relation provides a very good description of the data in the inset of 
Fig.\ \ref{fig:sigc_scaling}.
Beyond the characteristic system size $N_c$ both quantities $\left<\varepsilon_c\right>(N)$ and
$\left<\sigma_c\right>(N)$ are described by the same size scaling exponent $\alpha=2/3$.
The result shows that for small system sizes the macroscopic
strength of the bundle is determined by the extreme order statistics of the strength 
of single fibers, while above a characteristic system size this behavior 
breaks down and the average collective behavior of fibers of the bundle 
dominates \cite{hansen2015fiber}. 

\section{Discussion}
We investigated the effect of fat-tailed microscopic disorder on the macroscopic 
fracture strength of heterogeneous materials in the framework of a fiber bundle 
model with equal load sharing. The amount of disorder was controlled by 
varying the upper cutoff of fibers' strength and the power law exponent of the 
strength distribution. Analyzing the constitutive response of the system, we 
determined its phase diagram on the plane of control parameters: for law
values of the upper cutoff the bundle behaves in a completely brittle way where 
the breaking of the weakest fiber triggers the sudden collapse of the bundle.
For sufficiently high disorder a quasi-brittle response is obtained where macroscopic
failure is approached through stable cracking.

We focused on the size scaling of macroscopic strength of the bundle in the quasi-brittle
phase. Computer simulations revealed an astonishing size effect, i.e.\ for small system sizes
the bundle strength increases with the number of fibers such that the usual decreasing 
behavior sets on only above a characteristic system size. Fat-tailed disorder has the 
consequence that already at small system sizes strong fibers are included in the bundle
with a high probability. It implies that even a single fiber may be able
to keep the total load put on the system so that for small system sizes the macroscopic
bundle strength is determined by the extreme order statistics of the strength of single 
fibers. Since the fiber strength is bounded from above, for large enough system sizes 
the strongest fiber cannot compete with the load kept by the weaker fibers so that 
the regular decreasing size scaling gets restored. Based on this argument we could 
give an analytic description of the size scaling of bundles strength in the presence of 
fat-tailed disorder and determined the crossover system size, as well.

A similar strengthening behaviour for small system sizes has been observed 
in time dependent fiber bundles under localized load sharing conditions 
\cite{PhysRevE.63.021507,PhysRevE.80.066115}. It was found that for sufficiently small
power law exponents of the life consumption function of fibers a few long lived fibers 
dominate giving rise to an increased lifetime of the entire bundle. Consequently, the 
bundle lifetime is described by an extreme value distribution 
\cite{PhysRevE.63.021507,PhysRevE.80.066115}.

Our study is focused on the ELS limit of FBMs where all fibers keep the same load. 
In fibrous materials stress fluctuations natually arise due to the localized 
load sharing after fiber failures. In the strengthening regime, where 
extreme order statistics dominates the size scaling, no difference of ELS and 
LLS systems is expected. However, the crossover system size $N_c$ can depend on the 
precise form of the load redistribution scheme. Simulation studies in this direction
are in progress.

Recently, it has been shown that controlling the micro-structure 
\cite{shekhawat_toughness_2016} or the 
micro-scale disorder \cite{Miskin05012016} of materials novel type of materials 
can be taylored with desired properties for specialized applications. The scaling 
regime where the macroscopic strength increases with the system size may have potential
for materials design in future applications.

\begin{acknowledgments}
This work was supported through the New National Excellence Program of the Ministry 
of Human Capacities of Hungary.
We acknowledge the support of the project K119967 of NKFIH.
\end{acknowledgments}

\bibliography{/home/feri/papers/statphys_fracture}

\begin{thebibliography}{34}%
\makeatletter
\providecommand \@ifxundefined [1]{%
 \@ifx{#1\undefined}
}%
\providecommand \@ifnum [1]{%
 \ifnum #1\expandafter \@firstoftwo
 \else \expandafter \@secondoftwo
 \fi
}%
\providecommand \@ifx [1]{%
 \ifx #1\expandafter \@firstoftwo
 \else \expandafter \@secondoftwo
 \fi
}%
\providecommand \natexlab [1]{#1}%
\providecommand \enquote  [1]{``#1''}%
\providecommand \bibnamefont  [1]{#1}%
\providecommand \bibfnamefont [1]{#1}%
\providecommand \citenamefont [1]{#1}%
\providecommand \href@noop [0]{\@secondoftwo}%
\providecommand \href [0]{\begingroup \@sanitize@url \@href}%
\providecommand \@href[1]{\@@startlink{#1}\@@href}%
\providecommand \@@href[1]{\endgroup#1\@@endlink}%
\providecommand \@sanitize@url [0]{\catcode `\\12\catcode `\$12\catcode
  `\&12\catcode `\#12\catcode `\^12\catcode `\_12\catcode `\%12\relax}%
\providecommand \@@startlink[1]{}%
\providecommand \@@endlink[0]{}%
\providecommand \url  [0]{\begingroup\@sanitize@url \@url }%
\providecommand \@url [1]{\endgroup\@href {#1}{\urlprefix }}%
\providecommand \urlprefix  [0]{URL }%
\providecommand \Eprint [0]{\href }%
\providecommand \doibase [0]{http://dx.doi.org/}%
\providecommand \selectlanguage [0]{\@gobble}%
\providecommand \bibinfo  [0]{\@secondoftwo}%
\providecommand \bibfield  [0]{\@secondoftwo}%
\providecommand \translation [1]{[#1]}%
\providecommand \BibitemOpen [0]{}%
\providecommand \bibitemStop [0]{}%
\providecommand \bibitemNoStop [0]{.\EOS\space}%
\providecommand \EOS [0]{\spacefactor3000\relax}%
\providecommand \BibitemShut  [1]{\csname bibitem#1\endcsname}%
\let\auto@bib@innerbib\@empty
\bibitem [{\citenamefont {Alava}\ \emph {et~al.}(2006)\citenamefont {Alava},
  \citenamefont {Nukala},\ and\ \citenamefont
  {Zapperi}}]{alava_statistical_2006}%
  \BibitemOpen
  \bibfield  {author} {\bibinfo {author} {\bibfnamefont {M.}~\bibnamefont
  {Alava}}, \bibinfo {author} {\bibfnamefont {P.~K.}\ \bibnamefont {Nukala}}, \
  and\ \bibinfo {author} {\bibfnamefont {S.}~\bibnamefont {Zapperi}},\
  }\href@noop {} {\bibfield  {journal} {\bibinfo  {journal} {Adv. Phys.}\
  }\textbf {\bibinfo {volume} {55}},\ \bibinfo {pages} {349–476} (\bibinfo
  {year} {2006})}\BibitemShut {NoStop}%
\bibitem [{\citenamefont {Alava}\ \emph {et~al.}(2008)\citenamefont {Alava},
  \citenamefont {Nukala},\ and\ \citenamefont {Zapperi}}]{alava_role_2008}%
  \BibitemOpen
  \bibfield  {author} {\bibinfo {author} {\bibfnamefont {M.~J.}\ \bibnamefont
  {Alava}}, \bibinfo {author} {\bibfnamefont {P.~K. V.~V.}\ \bibnamefont
  {Nukala}}, \ and\ \bibinfo {author} {\bibfnamefont {S.}~\bibnamefont
  {Zapperi}},\ }\href@noop {} {\bibfield  {journal} {\bibinfo  {journal} {Phys.
  Rev. Lett.}\ }\textbf {\bibinfo {volume} {100}},\ \bibinfo {pages} {055502}
  (\bibinfo {year} {2008})}\BibitemShut {NoStop}%
\bibitem [{\citenamefont {Alava}\ \emph {et~al.}(2009)\citenamefont {Alava},
  \citenamefont {Nukala},\ and\ \citenamefont {Zapperi}}]{alava_size_2009}%
  \BibitemOpen
  \bibfield  {author} {\bibinfo {author} {\bibfnamefont {M.~J.}\ \bibnamefont
  {Alava}}, \bibinfo {author} {\bibfnamefont {P.~K. V.~V.}\ \bibnamefont
  {Nukala}}, \ and\ \bibinfo {author} {\bibfnamefont {S.}~\bibnamefont
  {Zapperi}},\ }\href@noop {} {\bibfield  {journal} {\bibinfo  {journal} {J.
  Phys. D: Appl. Phys.}\ }\textbf {\bibinfo {volume} {42}},\ \bibinfo {pages}
  {214012} (\bibinfo {year} {2009})}\BibitemShut {NoStop}%
\bibitem [{\citenamefont {Yamamoto}\ \emph {et~al.}(2011)\citenamefont
  {Yamamoto}, \citenamefont {Kun},\ and\ \citenamefont
  {Yukawa}}]{yamamoto_PhysRevE.83.066108}%
  \BibitemOpen
  \bibfield  {author} {\bibinfo {author} {\bibfnamefont {A.}~\bibnamefont
  {Yamamoto}}, \bibinfo {author} {\bibfnamefont {F.}~\bibnamefont {Kun}}, \
  and\ \bibinfo {author} {\bibfnamefont {S.}~\bibnamefont {Yukawa}},\
  }\href@noop {} {\bibfield  {journal} {\bibinfo  {journal} {Phys. Rev. E}\
  }\textbf {\bibinfo {volume} {83}},\ \bibinfo {pages} {066108} (\bibinfo
  {year} {2011})}\BibitemShut {NoStop}%
\bibitem [{\citenamefont {Galambos}(1978)}]{galambos_asymptotic_1978}%
  \BibitemOpen
  \bibfield  {author} {\bibinfo {author} {\bibfnamefont {J.}~\bibnamefont
  {Galambos}},\ }\href@noop {} {\emph {\bibinfo {title} {The Asymptotic Theory
  of Extreme Order Statistics}}}\ (\bibinfo  {publisher} {Wiley},\ \bibinfo
  {address} {New York},\ \bibinfo {year} {1978})\BibitemShut {NoStop}%
\bibitem [{\citenamefont {Weibull}(1939)}]{weibull_statistical_1939}%
  \BibitemOpen
  \bibfield  {author} {\bibinfo {author} {\bibfnamefont {W.}~\bibnamefont
  {Weibull}},\ }\href@noop {} {\emph {\bibinfo {title} {A statistical theory of
  the strength of materials}}}\ (\bibinfo  {publisher} {Generalstabens
  litografiska anstalt f\"orlag, Stockholm},\ \bibinfo {year}
  {1939})\BibitemShut {NoStop}%
\bibitem [{\citenamefont {Bazant}\ and\ \citenamefont
  {Planas}(1997)}]{bazant_fracture_1997}%
  \BibitemOpen
  \bibfield  {author} {\bibinfo {author} {\bibfnamefont {Z.~P.}\ \bibnamefont
  {Bazant}}\ and\ \bibinfo {author} {\bibfnamefont {J.}~\bibnamefont
  {Planas}},\ }\href@noop {} {\emph {\bibinfo {title} {Fracture and size effect
  in concrete and other quasibrittle materials}}}\ (\bibinfo  {publisher}
  {{CRC} Press, Boca Raton, {FL.}},\ \bibinfo {year} {1997})\BibitemShut
  {NoStop}%
\bibitem [{\citenamefont {Herrmann}\ \emph {et~al.}(1989)\citenamefont
  {Herrmann}, \citenamefont {Hansen},\ and\ \citenamefont
  {Roux}}]{herrmann_fracture_1989}%
  \BibitemOpen
  \bibfield  {author} {\bibinfo {author} {\bibfnamefont {H.~J.}\ \bibnamefont
  {Herrmann}}, \bibinfo {author} {\bibfnamefont {A.}~\bibnamefont {Hansen}}, \
  and\ \bibinfo {author} {\bibfnamefont {S.}~\bibnamefont {Roux}},\ }\href@noop
  {} {\bibfield  {journal} {\bibinfo  {journal} {Phys. Rev. B}\ }\textbf
  {\bibinfo {volume} {39}},\ \bibinfo {pages} {637–647} (\bibinfo {year}
  {1989})}\BibitemShut {NoStop}%
\bibitem [{\citenamefont {Hansen}\ \emph {et~al.}(1989)\citenamefont {Hansen},
  \citenamefont {Roux},\ and\ \citenamefont {Herrmann}}]{hansen_rupture_1989}%
  \BibitemOpen
  \bibfield  {author} {\bibinfo {author} {\bibfnamefont {A.}~\bibnamefont
  {Hansen}}, \bibinfo {author} {\bibfnamefont {S.}~\bibnamefont {Roux}}, \ and\
  \bibinfo {author} {\bibfnamefont {H.~J.}\ \bibnamefont {Herrmann}},\
  }\href@noop {} {\bibfield  {journal} {\bibinfo  {journal} {J. Phys. France}\
  }\textbf {\bibinfo {volume} {50}},\ \bibinfo {pages} {733} (\bibinfo {year}
  {1989})}\BibitemShut {NoStop}%
\bibitem [{\citenamefont {Batrouni}\ and\ \citenamefont
  {Hansen}(1998)}]{batrouni_fracture_1998}%
  \BibitemOpen
  \bibfield  {author} {\bibinfo {author} {\bibfnamefont {G.~G.}\ \bibnamefont
  {Batrouni}}\ and\ \bibinfo {author} {\bibfnamefont {A.}~\bibnamefont
  {Hansen}},\ }\href@noop {} {\bibfield  {journal} {\bibinfo  {journal} {Phys.
  Rev. Lett.}\ }\textbf {\bibinfo {volume} {80}},\ \bibinfo {pages} {325}
  (\bibinfo {year} {1998})}\BibitemShut {NoStop}%
\bibitem [{\citenamefont {Nukala}\ \emph {et~al.}(2004)\citenamefont {Nukala},
  \citenamefont {Simunovic},\ and\ \citenamefont
  {Zapperi}}]{nukala_percolation_2004}%
  \BibitemOpen
  \bibfield  {author} {\bibinfo {author} {\bibfnamefont {P.~V.~V.}\
  \bibnamefont {Nukala}}, \bibinfo {author} {\bibfnamefont {S.}~\bibnamefont
  {Simunovic}}, \ and\ \bibinfo {author} {\bibfnamefont {S.}~\bibnamefont
  {Zapperi}},\ }\href@noop {} {\bibfield  {journal} {\bibinfo  {journal} {J.
  Stat. Mech: Theor. Exp.}\ ,\ \bibinfo {pages} {P08001}} (\bibinfo {year}
  {2004})}\BibitemShut {NoStop}%
\bibitem [{\citenamefont {Bertalan}\ \emph {et~al.}(2014)\citenamefont
  {Bertalan}, \citenamefont {Shekhawat}, \citenamefont {Sethna},\ and\
  \citenamefont {Zapperi}}]{zapperi_PhysRevApplied.2.034008}%
  \BibitemOpen
  \bibfield  {author} {\bibinfo {author} {\bibfnamefont {Z.}~\bibnamefont
  {Bertalan}}, \bibinfo {author} {\bibfnamefont {A.}~\bibnamefont {Shekhawat}},
  \bibinfo {author} {\bibfnamefont {J.}~\bibnamefont {Sethna}}, \ and\ \bibinfo
  {author} {\bibfnamefont {S.}~\bibnamefont {Zapperi}},\ }\href@noop {}
  {\bibfield  {journal} {\bibinfo  {journal} {Phys. Rev. Applied}\ }\textbf
  {\bibinfo {volume} {2}},\ \bibinfo {pages} {034008} (\bibinfo {year}
  {2014})}\BibitemShut {NoStop}%
\bibitem [{\citenamefont {de~Arcangelis}\ \emph {et~al.}(1989)\citenamefont
  {de~Arcangelis}, \citenamefont {Hansen}, \citenamefont {Herrmann},\ and\
  \citenamefont {Roux}}]{de_arcangelis_scaling_1989}%
  \BibitemOpen
  \bibfield  {author} {\bibinfo {author} {\bibfnamefont {L.}~\bibnamefont
  {de~Arcangelis}}, \bibinfo {author} {\bibfnamefont {A.}~\bibnamefont
  {Hansen}}, \bibinfo {author} {\bibfnamefont {H.~J.}\ \bibnamefont
  {Herrmann}}, \ and\ \bibinfo {author} {\bibfnamefont {S.}~\bibnamefont
  {Roux}},\ }\href@noop {} {\bibfield  {journal} {\bibinfo  {journal} {Phys.
  Rev. B}\ }\textbf {\bibinfo {volume} {40}},\ \bibinfo {pages} {877} (\bibinfo
  {year} {1989})}\BibitemShut {NoStop}%
\bibitem [{\citenamefont {Andersen}\ \emph {et~al.}(1997)\citenamefont
  {Andersen}, \citenamefont {Sornette},\ and\ \citenamefont
  {Leung}}]{andersen_tricritical_1997}%
  \BibitemOpen
  \bibfield  {author} {\bibinfo {author} {\bibfnamefont {J.~V.}\ \bibnamefont
  {Andersen}}, \bibinfo {author} {\bibfnamefont {D.}~\bibnamefont {Sornette}},
  \ and\ \bibinfo {author} {\bibfnamefont {K.}~\bibnamefont {Leung}},\
  }\href@noop {} {\bibfield  {journal} {\bibinfo  {journal} {Phys. Rev. Lett.}\
  }\textbf {\bibinfo {volume} {78}},\ \bibinfo {pages} {2140–2143} (\bibinfo
  {year} {1997})}\BibitemShut {NoStop}%
\bibitem [{\citenamefont {Hansen}\ \emph {et~al.}(2015)\citenamefont {Hansen},
  \citenamefont {Hemmer},\ and\ \citenamefont {Pradhan}}]{hansen2015fiber}%
  \BibitemOpen
  \bibfield  {author} {\bibinfo {author} {\bibfnamefont {A.}~\bibnamefont
  {Hansen}}, \bibinfo {author} {\bibfnamefont {P.}~\bibnamefont {Hemmer}}, \
  and\ \bibinfo {author} {\bibfnamefont {S.}~\bibnamefont {Pradhan}},\
  }\href@noop {} {\emph {\bibinfo {title} {The Fiber Bundle Model: Modeling
  Failure in Materials}}},\ Statistical Physics of Fracture and Breakdown\
  (\bibinfo  {publisher} {Wiley},\ \bibinfo {year} {2015})\BibitemShut
  {NoStop}%
\bibitem [{\citenamefont {Kloster}\ \emph {et~al.}(1997)\citenamefont
  {Kloster}, \citenamefont {Hansen},\ and\ \citenamefont
  {Hemmer}}]{kloster_burst_1997}%
  \BibitemOpen
  \bibfield  {author} {\bibinfo {author} {\bibfnamefont {M.}~\bibnamefont
  {Kloster}}, \bibinfo {author} {\bibfnamefont {A.}~\bibnamefont {Hansen}}, \
  and\ \bibinfo {author} {\bibfnamefont {P.~C.}\ \bibnamefont {Hemmer}},\
  }\href@noop {} {\bibfield  {journal} {\bibinfo  {journal} {Phys. Rev. E}\
  }\textbf {\bibinfo {volume} {56}},\ \bibinfo {pages} {2615–2625} (\bibinfo
  {year} {1997})}\BibitemShut {NoStop}%
\bibitem [{\citenamefont {Kun}\ \emph {et~al.}(2006)\citenamefont {Kun},
  \citenamefont {Raischel}, \citenamefont {Hidalgo},\ and\ \citenamefont
  {Herrmann}}]{kun_extensions_2006}%
  \BibitemOpen
  \bibfield  {author} {\bibinfo {author} {\bibfnamefont {F.}~\bibnamefont
  {Kun}}, \bibinfo {author} {\bibfnamefont {F.}~\bibnamefont {Raischel}},
  \bibinfo {author} {\bibfnamefont {R.~C.}\ \bibnamefont {Hidalgo}}, \ and\
  \bibinfo {author} {\bibfnamefont {H.~J.}\ \bibnamefont {Herrmann}},\ }in\
  \href@noop {} {\emph {\bibinfo {booktitle} {Modelling Critical and
  Catastrophic Phenomena in Geoscience: A Statistical Physics Approach}}},\
  \bibinfo {series and number} {Lecture Notes in Physics},\ \bibinfo {editor}
  {edited by\ \bibinfo {editor} {\bibfnamefont {P.}~\bibnamefont
  {Bhattacharyya}}\ and\ \bibinfo {editor} {\bibfnamefont {B.~K.}\ \bibnamefont
  {Chakrabarti}}}\ (\bibinfo  {publisher} {{Springer-Verlag} Berlin Heidelberg
  New York},\ \bibinfo {year} {2006})\ pp.\ \bibinfo {pages}
  {57--92}\BibitemShut {NoStop}%
\bibitem [{\citenamefont {Hidalgo}\ \emph {et~al.}(2009)\citenamefont
  {Hidalgo}, \citenamefont {Kun}, \citenamefont {Kov\'acs},\ and\ \citenamefont
  {Pagonabarraga}}]{hidalgo_avalanche_2009}%
  \BibitemOpen
  \bibfield  {author} {\bibinfo {author} {\bibfnamefont {R.~C.}\ \bibnamefont
  {Hidalgo}}, \bibinfo {author} {\bibfnamefont {F.}~\bibnamefont {Kun}},
  \bibinfo {author} {\bibfnamefont {K.}~\bibnamefont {Kov\'acs}}, \ and\
  \bibinfo {author} {\bibfnamefont {I.}~\bibnamefont {Pagonabarraga}},\
  }\href@noop {} {\bibfield  {journal} {\bibinfo  {journal} {Phys. Rev. E}\
  }\textbf {\bibinfo {volume} {80}},\ \bibinfo {pages} {051108} (\bibinfo
  {year} {2009})}\BibitemShut {NoStop}%
\bibitem [{\citenamefont {Smith}(1982)}]{smith_asymptotic_1982}%
  \BibitemOpen
  \bibfield  {author} {\bibinfo {author} {\bibfnamefont {R.~L.}\ \bibnamefont
  {Smith}},\ }\href@noop {} {\bibfield  {journal} {\bibinfo  {journal} {Ann.
  Probab.}\ }\textbf {\bibinfo {volume} {10}},\ \bibinfo {pages} {137}
  (\bibinfo {year} {1982})}\BibitemShut {NoStop}%
\bibitem [{\citenamefont {McCartney}\ and\ \citenamefont
  {Smith}(1983)}]{mccartney_statistical_1983}%
  \BibitemOpen
  \bibfield  {author} {\bibinfo {author} {\bibfnamefont {L.~N.}\ \bibnamefont
  {McCartney}}\ and\ \bibinfo {author} {\bibfnamefont {R.~L.}\ \bibnamefont
  {Smith}},\ }\href {\doibase 10.1115/1.3167097} {\bibfield  {journal}
  {\bibinfo  {journal} {J. Appl. Mech}\ }\textbf {\bibinfo {volume} {50}},\
  \bibinfo {pages} {601} (\bibinfo {year} {1983})}\BibitemShut {NoStop}%
\bibitem [{\citenamefont {Harlow}(1985)}]{harlow_pure_1985}%
  \BibitemOpen
  \bibfield  {author} {\bibinfo {author} {\bibfnamefont {D.~G.}\ \bibnamefont
  {Harlow}},\ }\href@noop {} {\bibfield  {journal} {\bibinfo  {journal} {Proc.
  R. Soc. Lond. A}\ }\textbf {\bibinfo {volume} {397}},\ \bibinfo {pages} {211}
  (\bibinfo {year} {1985})}\BibitemShut {NoStop}%
\bibitem [{\citenamefont {Hansen}\ and\ \citenamefont
  {Hemmer}(1994)}]{hansen_burst_1994}%
  \BibitemOpen
  \bibfield  {author} {\bibinfo {author} {\bibfnamefont {A.}~\bibnamefont
  {Hansen}}\ and\ \bibinfo {author} {\bibfnamefont {P.~C.}\ \bibnamefont
  {Hemmer}},\ }\href@noop {} {\bibfield  {journal} {\bibinfo  {journal} {Phys.
  Lett. A}\ }\textbf {\bibinfo {volume} {184}},\ \bibinfo {pages} {394–396}
  (\bibinfo {year} {1994})}\BibitemShut {NoStop}%
\bibitem [{\citenamefont {Hidalgo}\ \emph {et~al.}(2002)\citenamefont
  {Hidalgo}, \citenamefont {Moreno}, \citenamefont {Kun},\ and\ \citenamefont
  {Herrmann}}]{hidalgo_fracture_2002}%
  \BibitemOpen
  \bibfield  {author} {\bibinfo {author} {\bibfnamefont {R.~C.}\ \bibnamefont
  {Hidalgo}}, \bibinfo {author} {\bibfnamefont {Y.}~\bibnamefont {Moreno}},
  \bibinfo {author} {\bibfnamefont {F.}~\bibnamefont {Kun}}, \ and\ \bibinfo
  {author} {\bibfnamefont {H.~J.}\ \bibnamefont {Herrmann}},\ }\href@noop {}
  {\bibfield  {journal} {\bibinfo  {journal} {Phys. Rev. E}\ }\textbf {\bibinfo
  {volume} {65}},\ \bibinfo {pages} {046148} (\bibinfo {year}
  {2002})}\BibitemShut {NoStop}%
\bibitem [{\citenamefont {{Dill-Langer}}\ \emph {et~al.}(2003)\citenamefont
  {{Dill-Langer}}, \citenamefont {Hidalgo}, \citenamefont {Kun}, \citenamefont
  {Moreno}, \citenamefont {Aicher},\ and\ \citenamefont
  {Herrmann}}]{dill-langer_size_2003}%
  \BibitemOpen
  \bibfield  {author} {\bibinfo {author} {\bibfnamefont {G.}~\bibnamefont
  {{Dill-Langer}}}, \bibinfo {author} {\bibfnamefont {R.~C.}\ \bibnamefont
  {Hidalgo}}, \bibinfo {author} {\bibfnamefont {F.}~\bibnamefont {Kun}},
  \bibinfo {author} {\bibfnamefont {Y.}~\bibnamefont {Moreno}}, \bibinfo
  {author} {\bibfnamefont {S.}~\bibnamefont {Aicher}}, \ and\ \bibinfo {author}
  {\bibfnamefont {H.~J.}\ \bibnamefont {Herrmann}},\ }\href@noop {} {\bibfield
  {journal} {\bibinfo  {journal} {Physica A}\ }\textbf {\bibinfo {volume}
  {325}},\ \bibinfo {pages} {547–560} (\bibinfo {year} {2003})}\BibitemShut
  {NoStop}%
\bibitem [{\citenamefont {Yewande}\ \emph {et~al.}(2003)\citenamefont
  {Yewande}, \citenamefont {Moreno}, \citenamefont {Kun}, \citenamefont
  {Hidalgo},\ and\ \citenamefont {Herrmann}}]{yewande_time_2003}%
  \BibitemOpen
  \bibfield  {author} {\bibinfo {author} {\bibfnamefont {O.~E.}\ \bibnamefont
  {Yewande}}, \bibinfo {author} {\bibfnamefont {Y.}~\bibnamefont {Moreno}},
  \bibinfo {author} {\bibfnamefont {F.}~\bibnamefont {Kun}}, \bibinfo {author}
  {\bibfnamefont {R.~C.}\ \bibnamefont {Hidalgo}}, \ and\ \bibinfo {author}
  {\bibfnamefont {H.~J.}\ \bibnamefont {Herrmann}},\ }\href@noop {} {\bibfield
  {journal} {\bibinfo  {journal} {Phys. Rev. E}\ }\textbf {\bibinfo {volume}
  {68}},\ \bibinfo {pages} {026116} (\bibinfo {year} {2003})}\BibitemShut
  {NoStop}%
\bibitem [{\citenamefont {Bazant}\ and\ \citenamefont
  {Pang}(2007)}]{bazant_activation_2007}%
  \BibitemOpen
  \bibfield  {author} {\bibinfo {author} {\bibfnamefont {Z.~P.}\ \bibnamefont
  {Bazant}}\ and\ \bibinfo {author} {\bibfnamefont {S.}~\bibnamefont {Pang}},\
  }\href@noop {} {\bibfield  {journal} {\bibinfo  {journal} {J. Mech. Phys.
  Solids}\ }\textbf {\bibinfo {volume} {55}},\ \bibinfo {pages} {91} (\bibinfo
  {year} {2007})}\BibitemShut {NoStop}%
\bibitem [{\citenamefont {Lehmann}\ and\ \citenamefont
  {Bernasconi}(2010)}]{lehmann_breakdown_2010}%
  \BibitemOpen
  \bibfield  {author} {\bibinfo {author} {\bibfnamefont {J.}~\bibnamefont
  {Lehmann}}\ and\ \bibinfo {author} {\bibfnamefont {J.}~\bibnamefont
  {Bernasconi}},\ }\href@noop {} {\bibfield  {journal} {\bibinfo  {journal}
  {Chem. Phys.}\ }\textbf {\bibinfo {volume} {375}},\ \bibinfo {pages} {591}
  (\bibinfo {year} {2010})}\BibitemShut {NoStop}%
\bibitem [{\citenamefont {Pradhan}\ \emph {et~al.}(2010)\citenamefont
  {Pradhan}, \citenamefont {Hansen},\ and\ \citenamefont
  {Chakrabarti}}]{pradhan_failure_2010}%
  \BibitemOpen
  \bibfield  {author} {\bibinfo {author} {\bibfnamefont {S.}~\bibnamefont
  {Pradhan}}, \bibinfo {author} {\bibfnamefont {A.}~\bibnamefont {Hansen}}, \
  and\ \bibinfo {author} {\bibfnamefont {B.~K.}\ \bibnamefont {Chakrabarti}},\
  }\href@noop {} {\bibfield  {journal} {\bibinfo  {journal} {Rev. Mod. Phys.}\
  }\textbf {\bibinfo {volume} {82}},\ \bibinfo {pages} {499} (\bibinfo {year}
  {2010})}\BibitemShut {NoStop}%
\bibitem [{\citenamefont {Newman}\ and\ \citenamefont
  {Gabrielov}(1991)}]{newman_failure_1991}%
  \BibitemOpen
  \bibfield  {author} {\bibinfo {author} {\bibfnamefont {W.~I.}\ \bibnamefont
  {Newman}}\ and\ \bibinfo {author} {\bibfnamefont {A.~M.}\ \bibnamefont
  {Gabrielov}},\ }\href@noop {} {\bibfield  {journal} {\bibinfo  {journal}
  {Int. J. Fract.}\ }\textbf {\bibinfo {volume} {50}},\ \bibinfo {pages}
  {1–14} (\bibinfo {year} {1991})}\BibitemShut {NoStop}%
\bibitem [{\citenamefont {Danku}\ and\ \citenamefont
  {Kun}(2016)}]{danku_disorder_2016}%
  \BibitemOpen
  \bibfield  {author} {\bibinfo {author} {\bibfnamefont {Z.}~\bibnamefont
  {Danku}}\ and\ \bibinfo {author} {\bibfnamefont {F.}~\bibnamefont {Kun}},\
  }\href@noop {} {\bibfield  {journal} {\bibinfo  {journal} {J. Stat. Mech.:
  Theor. Exp.}\ }\textbf {\bibinfo {volume} {2016}},\ \bibinfo {pages} {073211}
  (\bibinfo {year} {2016})}\BibitemShut {NoStop}%
\bibitem [{\citenamefont {Newman}\ and\ \citenamefont
  {Phoenix}(2001)}]{PhysRevE.63.021507}%
  \BibitemOpen
  \bibfield  {author} {\bibinfo {author} {\bibfnamefont {W.~I.}\ \bibnamefont
  {Newman}}\ and\ \bibinfo {author} {\bibfnamefont {S.~L.}\ \bibnamefont
  {Phoenix}},\ }\href@noop {} {\bibfield  {journal} {\bibinfo  {journal} {Phys.
  Rev. E}\ }\textbf {\bibinfo {volume} {63}},\ \bibinfo {pages} {021507}
  (\bibinfo {year} {2001})}\BibitemShut {NoStop}%
\bibitem [{\citenamefont {Phoenix}\ and\ \citenamefont
  {Newman}(2009)}]{PhysRevE.80.066115}%
  \BibitemOpen
  \bibfield  {author} {\bibinfo {author} {\bibfnamefont {S.~L.}\ \bibnamefont
  {Phoenix}}\ and\ \bibinfo {author} {\bibfnamefont {W.~I.}\ \bibnamefont
  {Newman}},\ }\href@noop {} {\bibfield  {journal} {\bibinfo  {journal} {Phys.
  Rev. E}\ }\textbf {\bibinfo {volume} {80}},\ \bibinfo {pages} {066115}
  (\bibinfo {year} {2009})}\BibitemShut {NoStop}%
\bibitem [{\citenamefont {Shekhawat}(2016)}]{shekhawat_toughness_2016}%
  \BibitemOpen
  \bibfield  {author} {\bibinfo {author} {\bibfnamefont {A.}~\bibnamefont
  {Shekhawat}},\ }\href {http://arxiv.org/abs/1611.01719} {\bibfield  {journal}
  {\bibinfo  {journal} {arXiv:1611.01719 [cond-mat]}\ } (\bibinfo {year}
  {2016})}\BibitemShut {NoStop}%
\bibitem [{\citenamefont {Miskin}\ \emph {et~al.}(2016)\citenamefont {Miskin},
  \citenamefont {Khaira}, \citenamefont {de~Pablo},\ and\ \citenamefont
  {Jaeger}}]{Miskin05012016}%
  \BibitemOpen
  \bibfield  {author} {\bibinfo {author} {\bibfnamefont {M.~Z.}\ \bibnamefont
  {Miskin}}, \bibinfo {author} {\bibfnamefont {G.}~\bibnamefont {Khaira}},
  \bibinfo {author} {\bibfnamefont {J.~J.}\ \bibnamefont {de~Pablo}}, \ and\
  \bibinfo {author} {\bibfnamefont {H.~M.}\ \bibnamefont {Jaeger}},\
  }\href@noop {} {\bibfield  {journal} {\bibinfo  {journal} {Proc. Nat. Acad.
  Sci.}\ }\textbf {\bibinfo {volume} {113}},\ \bibinfo {pages} {34} (\bibinfo
  {year} {2016})}\BibitemShut {NoStop}%
\end{thebibliography}%

\end{document}